\pdfoutput=1
\documentclass[10pt, doublecolumn]{IEEEtran}
\IEEEoverridecommandlockouts
\usepackage{cite}
\usepackage{indentfirst}
\usepackage{graphicx}
\usepackage{changepage}
\usepackage{stfloats}
\usepackage{amsfonts,amssymb}
\usepackage{bm}
\usepackage{algorithm}
\usepackage{algorithmic}
\usepackage{amsmath}
\usepackage{amsmath,amssymb,amsfonts}
\usepackage{times}
\usepackage{url}
\usepackage{cite}
\usepackage{textcomp}
\usepackage{xcolor}
\usepackage{color}
\usepackage{setspace}
\usepackage{enumitem}
\usepackage{float}
\usepackage{diagbox}
\usepackage[T1]{fontenc}
\usepackage[utf8]{inputenc}
\usepackage{authblk}
\usepackage{threeparttable}
\usepackage{booktabs}
\usepackage{footnote}
\usepackage{graphicx}
\usepackage{pifont}
\usepackage{subfigure}
\usepackage{mathrsfs}
\usepackage{makecell}
\usepackage{array}
\usepackage{booktabs}
\usepackage{lipsum}
\usepackage{multirow}
\usepackage{adjustbox}
\usepackage{bbding}
\usepackage{tabularx}
\usepackage{colortbl}
\usepackage{threeparttable}
\begin{document}
\title{ Integrated Sensing and Communication Meets Smart Propagation Engineering: Opportunities and Challenges}

\author{
	Kaitao Meng, \textit{Member, IEEE}, Christos Masouros, \textit{Fellow, IEEE}, Kai-Kit Wong, \textit{Fellow, IEEE}, Athina P. Petropulu, \textit{Fellow, IEEE}, and Lajos Hanzo, \textit{Fellow, IEEE}
	\thanks{{K. Meng, C. Masouros, and K. Wong are with the Department of Electronic and Electrical Engineering, University College London, London, UK }{(emails: \{kaitao.meng, c.masouros, kai-kit.wong\}@ucl.ac.uk).} A. P. Petropulu is with the Department of Electrical and Computer Engineering, Rutgers University, Piscataway, NJ 08901 USA (email: athinap@rustlers.edu). Lajos Hanzo is with the University of Southampton. (email: lh@ecs.soton.ac.uk) }
}

\maketitle

\begin{abstract}
	Both smart propagation engineering as well as integrated sensing and communication (ISAC) constitute promising candidates for next-generation (NG) mobile networks. We provide a synergistic view of these technologies, and explore their mutual benefits. First, moving beyond just intelligent surfaces, we provide a holistic view of the engineering aspects of smart propagation environments. By delving into the fundamental characteristics of intelligent surfaces, fluid antennas, and unmanned aerial vehicles, we reveal that more efficient control of the pathloss and fading can be achieved, thus facilitating intrinsic integration and mutual assistance between sensing and communication functionalities. In turn, with the exploitation of the sensing capabilities of ISAC to orchestrate the efficient configuration of radio environments, both the computational effort and signaling overheads can be reduced. We present indicative simulation results, which verify that cooperative smart propagation environment design significantly enhances the ISAC performance. Finally, some promising directions are outlined for combining ISAC with smart propagation engineering. 
\end{abstract}


\section{Introduction}
\par
As a promising candidate technology for next-generation (NG) networks, integrated sensing and communication (ISAC) relies on a unified wireless infrastructure and its spectral resources to convey the desired information, while exploiting echo signals for sensing. As such, ISAC seamlessly delivers sensing and communication (S\&C) services in a timely, energy-, and spectrally efficient manner \cite{Zhang2021Overview}. With the advances of multiple-input multiple-output (MIMO) and millimetre wave (mmWave)/terahertz (THz) technologies, ISAC is expected to provide high-throughput, ultra-reliable, and low-latency wireless communications, as well as ultra-precise, high-resolution, and robust wireless sensing \cite{Liu2022Integrated}. Nevertheless, for conventional communication networks having stationary antennas, fixed base station (BS) topology, and random/uncontrolled channel fading, the performance of S\&C may be severely restricted by transmission blockages, excessive connectivity demands, and by the conflicting design objectives of ISAC networks \cite{Zhang2021Overview, Liu2022Integrated}. 

By utilizing state-of-the-art smart technologies like intelligent surfaces \cite{ElMossallamy2020Reconfigurable}, fluid antenna systems (FAS) \cite{Wong2021Fluid}, and holographic MIMO \cite{Huang2020Holographic}, combined with innovative cell-free and mobile air-ground networks \cite{Meng2023Throughput}, the propagation environments can be beneficially managed with increased flexibility and efficiency, resulting in enhanced ISAC performance. Towards this end, in this article, we attempt a step forward from previous studies of intelligent surfaces conceived for propagation control \cite{Liu2023Integrated}, by harnessing a holistic view of the latest channel engineering technologies. We reveal that by taking advantage of these novel antenna techniques and network architectures, mutual assistance between S\&C is improved, resulting in sensing-assisted communication and communication-assisted sensing. Furthermore, they also offer new opportunities to design cooperative ISAC networks for signal power enhancement and efficient interference mitigation.

While these smart technologies are capable of significantly enhancing the S\&C performance attained, their advantages come at the expense of high latency, overheads, and power consumption imposed by the control operations. The radio environment should be carefully managed according to the dynamic requirements of specific S\&C tasks, as excessive adaptation may result in increased signaling overheads and resource consumption. It is therefore necessary to allocate the resources in accordance with the dynamics of the environments to ensure substantial returns. We note that leveraging the sensing capability assists in understanding the locations of served/detected users/objects, increasing awareness of potential blockages, and reducing signaling overheads required for environmental control. Thus, ISAC technologies can in turn assist in beneficially ameliorating the propagation environments, hence resulting in a win-win integration with mutual benefits. However, achieving improved ISAC performance and more efficient propagation control is a challenging new problem.

Against these backdrop, we offer a comprehensive overview of the integration of smart propagation engineering and ISAC systems, pointing out key challenges, exploring potential solutions, and identifying open future directions. Section \ref{KeyTechnology} provides a more general definition for smart propagation engineering, including several enabling techniques. Section \ref{SectionSmartEnvironmentsForISAC} briefly discusses propagation engineering for ISAC, while Section \ref{SectionISACForSmartEnvironments} presents ISAC solutions for radio environmental control. Section \ref{Extensions} enlists compelling future directions in ISAC enhanced by smart propagation engineering. Section \ref{CaseStudy} presents simulation-based verification of smart propagation engineering for ISAC. Finally, Section \ref{Conclusions} concludes this work.

\section{Key technologies of Smart Environments}
\label{KeyTechnology}

In this section, we provide a holistic view of smart propagation engineering: They utilize various forms of antennas and network entities for flexibly controlling the propagation between transmitters and receivers with increased degrees of freedom (DoF), and achieve smart time/frequency/spatial domain resource allocation for various sensing, communication, and computing tasks. In the following, we explore several techniques with the ability of radio environmental control and discuss their characteristics, advantages, challenges, and promising future directions.

\begin{figure*}[t]
	\centering
	\includegraphics[width=13cm]{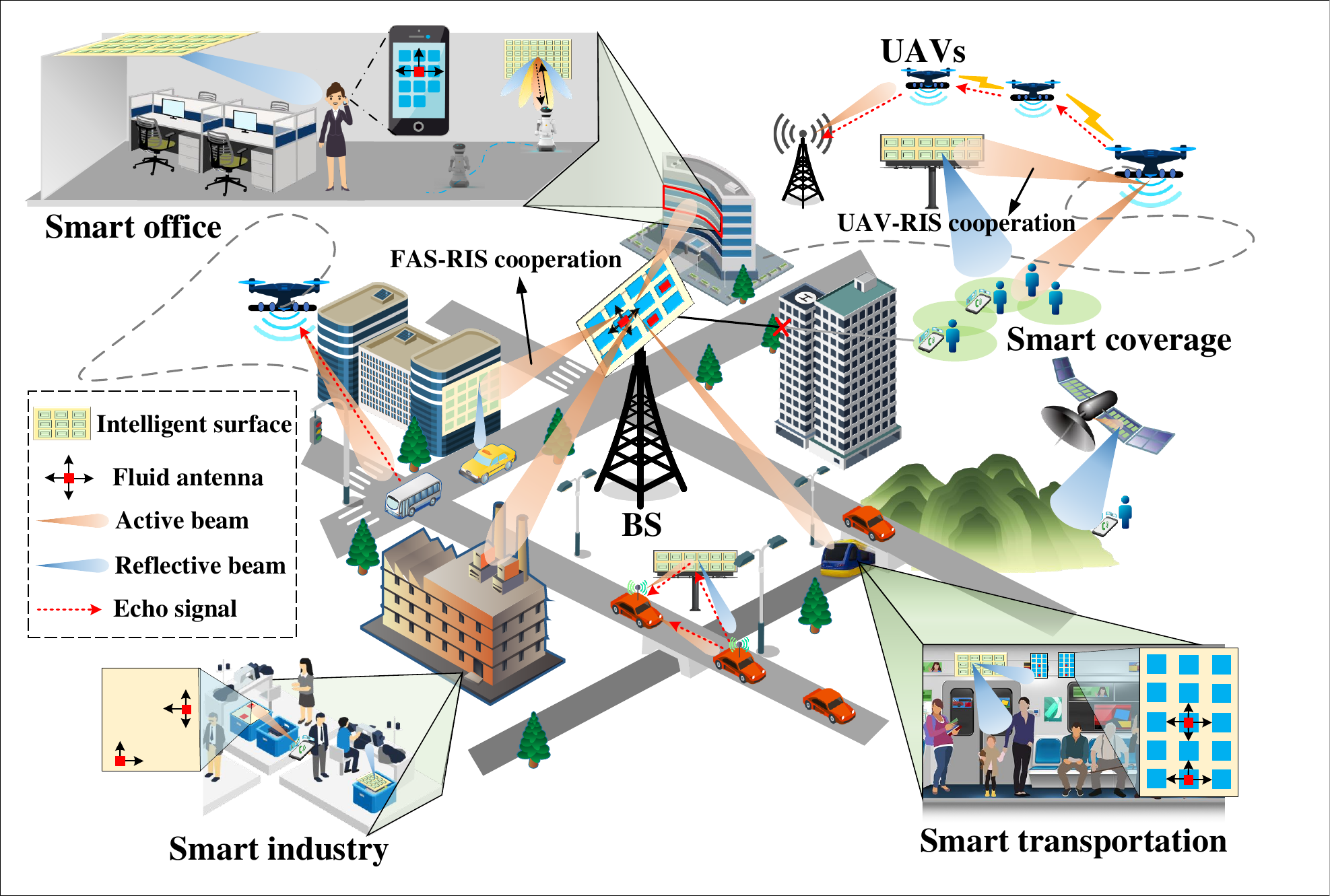}
	\vspace{-3mm}
	\caption{Illustration of Smart Environments for ISAC.}
	\label{figure1}
\end{figure*}

\subsection{Intelligent Surfaces}
With the exploitation of myriads of low-cost reflecting/refracting elements, reconfigurable intelligent surfaces (RISs) \cite{Meng2023Sensing} and stacked intelligent metasurfaces (SIMs) \cite{An2023Stacked} can adaptively ameliorate the propagation channel between transmitters and receivers by beneficially designing their phase shifts and/or amplification. Intelligent surfaces can be placed in strategic locations to circumvent blocked line-of-sight (LoS) connections between transmitters and receivers, as shown in Fig.~\ref{figure1}. Alongside this deployment strategy, intelligent surfaces can also be mounted onto mobile platforms, such as vehicles, drones, and even user terminals, providing a promising solution to enable these objects to modulate ambient signals as well as reflections, and actively assist their own localization and other services \cite{Meng2023Sensing}. With a larger surface operating in a higher frequency band, RISs often implicate operation in the near field, where the spherical nature of the electromagnetic waves has to be considered. This in turn offers new opportunities to exploit precise spatial beam focusing both in angle and range, and to design bent beams for avoiding blockages \cite{singh2023wavefront}. These offer profound opportunities to improve the propagation environment.

\subsection{Unmanned Aerial Vehicles (UAVs) and Other Autonomous Platforms}
For complementing conventional terrestrial cellular networks, BSs or relays may also be installed on unmanned aerial/ground/underwater vehicles to provide enhanced coverage, as shown in Fig.~\ref{figure1}. By designing the placements/trajectories of these mobile BSs/relays on demand, the network topology can be dynamically adjusted in accordance with specific network requirements and services. For example, by exploiting the UAVs' high mobility and strong air-ground LoS links, larger coverage, flexible observation position, and enhanced S\&C performance can be achieved \cite{Meng2023Throughput}. However, UAVs typically encounter challenges due to strict constraints on their size, weight, and power, which inevitably limit their capabilities. Additionally, stronger air-to-ground LoS links may cause severe interference. Besides these autonomous vehicles, high-altitude platforms (HAPs) are able to offer observation or communication services, typically powered with renewable sources and thus offering services with net-zero emissions. Moreover, thanks to the reduced launch cost and deployment time, low earth orbit (LEO) satellite systems offer promising opportunities for flexible global coverage.

\subsection{Flexible-Position FAS-aided Communications} 

In FAS \cite{Wong2021Fluid}, the positions of antennas at transmitters and/or receivers can be dynamically changed to obtain improved channel conditions, e.g., enhancing the channel gain and reducing the potential interference. A massive array of fluid antennas provides greater flexibility in designing beam width, avoiding undesirable side lobes, adjusting the frequency response of the array, and improving interference management. This is in contrast to traditional fixed-position antenna arrays, which suffer from reduced array gains due to inherent limitations in array geometry. Moreover, altering the positions of antennas could modify the Rayleigh distance \cite{Cui2023Near}, which sets the boundary between the far-field and near-field channels. While this further augments the near-field capabilities of RIS and the resultant propagation benefits, in practice, it is challenging to appropriately design the transmit beamforming vectors around the Rayleigh distance due to the dynamic antenna geometry. Moreover, existing far-field and near-field channel estimation schemes cannot be directly used for accurately estimating the "mixed"-field MIMO channel, which requires further investigation.

\begin{table*}[t] 
	\centering
	\caption{Illustration of several technologies roles in ISAC applications.}
	\label{Table2}
	\begin{tabular}{ |p l | l | l| l | l |}
		\hline
		\multicolumn{1}{|c|}{\makecell[l]{{\textbf{\!\! Smart Propagation}}\\ \textbf{Engineering technology$^\star$ \!\!\!\!\!}}} & {\makecell[c]{\textbf{Category }}}  & {\makecell[c]{\textbf{Main functionalities}}} & {\makecell[c]{{\textbf{Smart Propagation \!\!}}\\ \textbf{for ISAC}}} &  {\makecell[c]{{\textbf{ISAC for}}\\ \textbf{Smart Propagation \!\!}}}  \\ 
		\hline
		\multicolumn{1}{|l|}{\multirow{1}{*}{\begin{tabular}[l]{l}Intelligent surfaces\end{tabular}}}  & {\multirow{2}{*}{\begin{tabular}[c]{c}Adaptive channel\end{tabular}}} & {\makecell[c]{Control for}} & {\makecell[c]{Improve channel DoF,}} & {\makecell[c]{low-cost/latency}}  \\ 
		\multicolumn{1}{|c|}{(e.g., RIS, SIM)}   & {\makecell[c]{}}  & {\makecell[c]{ propagation channel }} & {\makecell[c]{enhance sensing diversity,}}  & {\makecell[c]{ channel estimation,}}  \\ 
		\cline{1-3}
		\multicolumn{1}{|c|}{\makecell[l]{UAVs, HAPs}}  & {\multirow{2}{*}{\begin{tabular}[c]{c}Mobile BS/relay\end{tabular}}} &{\makecell[c]{Management  for}}&{\makecell[c]{higher spectral efficiency,}}  & {\makecell[c]{blockage awareness}}  \\ 
		\cline{1-1}
		\multicolumn{1}{|c|}{\makecell[l]{LEO satellite}}    & {\makecell[c]{}} & {transmission topology} & {\makecell[c]{interference suppression, }} & {\makecell[c]{channel control,}}\\ 
		\cline{1-3}
		\multicolumn{1}{|c|}{\makecell[l]{Fluid antenna}} & {\multirow{2}{*}{\begin{tabular}[c]{@{}c@{}}Advanced antennas \end{tabular}}} &  {\makecell[c]{Transmitter/receiver }} & {\makecell[c]{S\&C coordination gain,}}  & {\makecell[c]{{AI-aided predictive}}}   \\ 
		\cline{1-1}
		\multicolumn{1}{|c|}{\makecell[c]{Holographic MIMO}}   & {}   &  {\makecell[c]{antenna geometry}}& {\makecell[c]{multi-cell cooperation}} & {\makecell[c]{resource allocation}} \\  
		\hline
		\noalign{\vskip 2mm}   
		\hline
	\end{tabular}
\begin{tablenotes}
	\footnotesize
	\item $^\star$By smart propagation engineering technology, we refer to any hardware technology that has the ability to engineer the radio propagation channel.
\end{tablenotes}
\end{table*} 
\subsection{Cooperative Smart Environment Control}

Effectively combining the above-mentioned techniques can proactively adapt to the wireless propagation environments. We categorize these smart technologies into three types based on their functions and objectives, as presented in Table \ref{Table2}. These technologies may cooperate to efficiently manage the radio environment in terms of time, energy, and cost by leveraging their shared features and differences. Typically, UAVs work hand-in-hand with RISs to create virtual LoS channels for improving S\&C coverage, as shown in the top-right corner of Fig.~\ref{figure1}. Consequently, the channel gain improvement enabled by the exploitation of RISs can eliminate the need for UAVs to fly closer to users or targets, thereby reducing the UAV's flight distance, power consumption and increasing its recharge interval. An example architecture was demonstrated in \cite{Nguyen2022UAV} where the authors proposed a UAV-aided RIS framework for maximizing the network throughput. On the other hand, FASs can adjust the antenna position according to the UAV trajectory and the RIS phase shifts, leading to more efficient exploitation of both random environmental scattering and controllable environmental reflection. In turn, RISs can be optimized concerning the antenna position for stronger signal power and for effective passive beam focusing. 

Jointly optimizing the trajectory of mobile BSs, the phase shifts of RIS elements, and the position of FASs is extremely challenging due to the coupling of variables. However, it also presents new opportunities for fully unleashing the power of smart propagation engineering. It is noted that more resource cooperation provides improved flexibility to manage the radio environment, but it inevitably introduces more interactions amongst devices and more complex channel estimation processes. Therefore, when controlling the radio environment based on the near-instantaneous radio map, an intriguing question is how to effectively harness these three types of resources for collaborative propagation engineering at low computational complexity and low signaling overhead. Moreover, to provide efficient, reliable, and robust S\&C services that meet the dynamic requirements of tasks, a cooperative smart propagation regime centered on users/targets remains an elusive challenge.

\section{Smart Propagation Engineering for ISAC}
\label{SectionSmartEnvironmentsForISAC}
In this section, we discuss opportunities for exploiting smart technologies to provide enhanced ISAC services by improving the S\&C channel gain, achieving augmented mutual assistance between S\&C, and ultimately for designing cooperative ISAC networks.

\subsection{Improving S\&C Performance via Enhancing the Channel Gain}
To achieve enhanced S\&C performance, a key challenge of ISAC assisted by smart propagation engineering is to simultaneously achieve the potentially conflicting channel control objectives of S\&C. Specifically, cooperative radio environmental control necessitates the creation of high-rank channels associated with greater flexibility for data transmission, thereby achieving improved spatial resource allocation for enhancing the multiplexing gain, diversity gain, and interference nulling. By contrast, for sensing, typically only LoS links are useful, and non-LoS (NLoS) links are treated as unfavorable interference \cite{Liu2023Integrated}. Thus, cooperative techniques should establish more LoS links for providing improved sensing coverage, more diverse observation angles, and richer target parameters.
Propagation engineering offers key opportunities in fine-tuning the propagation channel for striking a balance between these conflicting S\&C objectives of ISAC systems by optimizing the trajectory/placement of mobile transmitters/receivers, RIS phase shifts, FAS positions, etc. For instance, treated as both a mobile BS and a synthetic radar aperture, a UAV may optimize its trajectory to strike a balance between pathloss and channel gains for communication users, and observation diversity for targets. Moreover, to optimize the antenna position in FAS, smart propagation engineering may exploit channel paths to amplify the effective channel gains from the BS to all users. On the other hand, it additionally has to create larger antenna array apertures for capturing more echo signals from multiple directions, thereby improving the sensing performance attained.

\subsection{Enhanced Integration Gain Between S\&C}
An ISAC system can optimize its performance by strengthening the coupling between its S\&C channels \cite{chepuri2022integrated}. This facilitates more efficient exploitation of unified signals for S\&C tasks. In the following, we outline how smart propagation engineering enhances the correlation of S\&C channels. Firstly, considering that the S\&C channel correlation is determined by the angular separations between users and targets, the trajectory of mobile BSs significantly impacts the resource allocation and waveform design of mobile ISAC systems. Thus it is essential to jointly design the transmit beamforming and trajectory to enhance the S\&C performance \cite{Meng2023Throughput}. Secondly, the S\&C subspace of RISs having weak coupling may be rotated for improved coupling by adjusting the RIS phase shifts, hence striking improved S\&C tradeoffs \cite{chepuri2022integrated}. Thirdly, in FAS, the new DoF attained by antenna position optimization can be exploited for maximizing the S\&C signal power over a desired direction and for simultaneously reducing the potential interference. Evidently, the collaborative resource design of UAVs, RISs, and FASs offers new opportunities for improving the S\&C channel correlation. In particular, for operation in the near-field, achieving efficient cooperative control offers new opportunities for exploiting the channel correlation between users and targets not only in the angular domain but also in the distance domain \cite{Cui2023Near}.

\subsection{Smart Environment for Near-Field ISAC}

\begin{figure}[t]
	\centering
	\includegraphics[width=8cm]{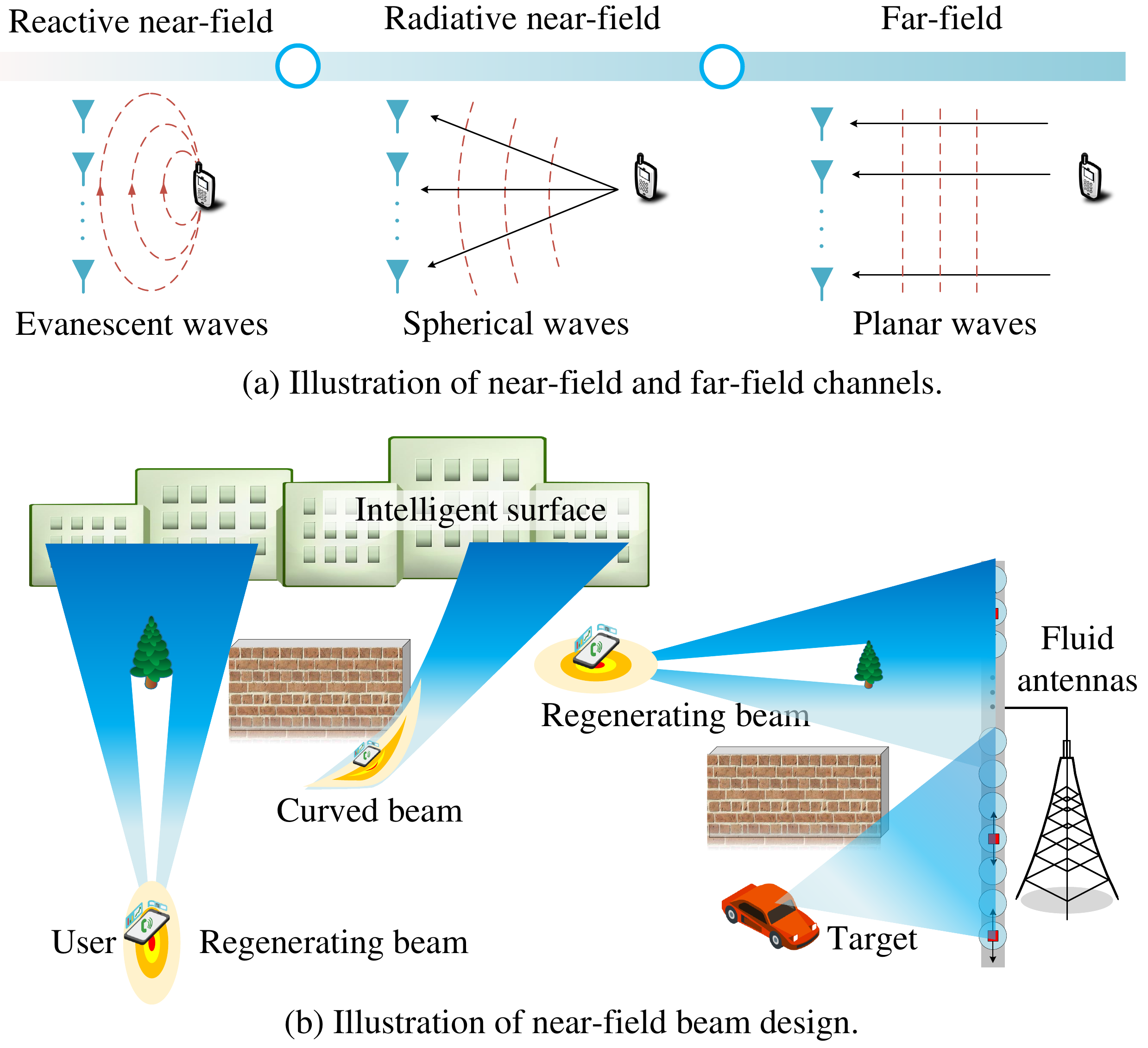}
	\vspace{-3mm}
	\caption{Illustration of ISAC for Near-field Smart Environments.}
	\label{figure4}
\end{figure}

Again, the electromagnetic (EM) field radiated from antennas can be partitioned into near-field and far-field regions, where the EM waves in these regions exhibit different propagation properties, as shown in Fig. \ref{figure4}(a). Specifically, for targets/users near the antenna array, smart propagation engineering should consider spherical wave modeling, as the planar wave approximation is no longer valid. In contrast to far-field beamforming design, a transmitter array is designed to concentrate the beam at a specific location, termed as beamfocusing. This enables the formation of spot beams (beam power concentrated on a specific angle and range), allowing for simultaneous serving/detection of users/targets in the same direction, which is not easily accomplished in the far field. Importantly, due to potential blockages, different sets of clusters/users/targets may be visible from different portions of the antenna array or intelligent surfaces, as shown in Fig.~\ref{figure4}(b). In this case, given the spherical wavefronts, the radiation from the unobstructed portion of the aperture can still converge to the desired focal point of beam focusing, as shown in Fig.~\ref{figure4}(b). Furthermore, the near field may even permit the generation of curved beams that circumvent obstacles \cite{singh2023wavefront}. For instance, the self-regenerating properties of Bessel beams \cite{singh2023wavefront} restore the beam after encountering an obstacle with negligible reduction in radiation intensity. These properties facilitate the provision of robust S\&C services in obstructed scenarios. In the near-field scenarios, variations of Doppler across these antennas offer new opportunities to estimate the radial and transverse velocities of targets, in turn also presenting new challenges to smart propagation engineering for ISAC systems. 



\subsection{Cooperative ISAC Networks}

For network-level ISAC, inter-cell interference is a critical issue that limits the overall network performance. Interference suppression via smart propagation engineering is a promising technique of improving the S\&C performance of ISAC networks, yet there is a paucity of related investigations in the literature. For instance, exploiting the UAV mobility combined with beamforming design effectively reduces the interference between BSs and UAVs. The RIS phase shift design should aim for reducing channel correlation between distinct users/targets and BSs, thereby mitigating multi-user/target interference \cite{ElMossallamy2020Reconfigurable}. FASs can exploit the potential diversity provided by the environment to avoid inter-cell S\&C interference. In addition to coordinated interference management, the cooperative S\&C benefits of ISAC networks is another promising paradigm. Specifically, by fully exploiting the potential of smart propagation engineering, a powerful combination of distributed/multistatic MIMO radar and coordinated multipoint transmission/reception can be supported. For instance, UAVs can serve as mobile transmitters/receivers to enhance S\&C coordination. Furthermore, intelligent surfaces play a crucial role in establishing additional S\&C channels, while FASs have the capability of dynamically adjusting the antenna geometry, for meeting near-instantaneous S\&C requirements. However, designing a cooperative scheme for the smart environments of ISAC networks that effectively balances performance against complexity remains a challenging open issue hinging on sophisticated resource allocation.

At the network level of ISAC, conflicting metrics such as coverage and quality of service have to be balanced to meet the demanding overall network performance requirements. Specifically, if larger clusters of BSs work together to provide improved S\&C services, the overall network spectrum efficiency will generally be reduced, since fewer users/targets can be served simultaneously. Furthermore, the challenge in designing and optimizing the resources of the ISAC network lies in judiciously balancing the various performance indicators, while taking into account both the channel fading and the unknown locations of targets/users. To tackle this issue, a useful approach is to apply stochastic geometry tools, which consider the random locations of BSs, users, and targets, along with random channel fading. This allows for modelling the relationships amongst various parameters and consequently facilitates more efficient optimization of the network resources, RIS/UAV positioning and FAS design.

\section{ISAC for Smart Environments}
\label{SectionISACForSmartEnvironments}
In this section, we explore the ISAC-benefits of smart propagation engineering, including cooperative engineering, low-overhead control, predictive resource allocation, and blockage awareness.

\subsection{ISAC for Cooperative Smart Environment}

Harnessing numerous UAVs, RISs, and FASs for effective cooperation is a promising option for improving the S\&C performance and for reducing the time/energy consumption, but it also poses challenges related to the essential information exchange across systems. To tackle this issue, the S\&C functionalities of ISAC systems can be exploited to facilitate the collaboration of the various system components. For instance, based on the dynamics of targets and users, it is conducive to combine sensing results and dispatch request messages for collaboration based on the fusion of sensory data \cite{Liu2022Integrated}. This enables these systems to operate in dynamic clusters, identifying potent cooperation that maximizes gains, while avoiding deficient cooperation having only incremental gains. Moreover, incorporating sensory data, such as the positions of users and targets, can prevent unnecessary control procedures. For example, if the channel fluctuation in a particular area drops below a certain threshold, the resource allocations do not have to be updated. It is worth noting that ISAC is capable of managing radio propagation across various geographical scales. Since the UAV movement is of limited speed, it is more suitable for propagation engineering in scenarios of more relaxed latency requirements. Intelligent surface technology by contrast allows for near-real-time applications, but over a limited area. With the use of an FAS, the antenna position can be adjusted based on the channel status and user mobility characteristics.

\subsection{ISAC for Low Overhead Control}
\label{SensingAssistedCommunication}

Resource optimization in smart UAV, RIS, and FAS aided scenarios relies heavily on the channel state information (CSI) as well as on the locations of users and targets. Estimating the CSI becomes complex as the number of nodes plus active and passive antennas increases, especially when these systems collaborate with each other. Therefore, sensing-assisted CSI estimation in ISAC systems shows great potential to reduce training overhead \cite{Meng2023Sensing, mura2023enhanced}. By representing CSI in terms of path-delay and angle in fewer dimensions, we can significantly reduce the CSI estimation overhead. Employing sensing for improved beam steering may even circumvent CSI acquisition altogether by exploiting the direction finding capabilities of radar to inform the BS of the user's direction for high-gain beam focusing. In \cite{Meng2023Sensing}, an ISAC framework using vehicle-mounted intelligent surfaces was conceived for strategically directing the echo signals towards the sensing receivers. By exploiting the high-accuracy sensing results, the signaling overhead of the RIS phase shift design may be effectively reduced. Moreover, the uplink data transmission of communication users may also be harnessed for gathering information on the environmental and target status for smart propagation engineering.

Different smart propagation engineering technologies achieve low-overhead control in unique ways. For instance, incorporating sensors into the RIS enhances its capability of analyzing echo signals. Hence, the sensing results support self-adaptive passive beamforming design, without necessitating any control signals from the BS. Sensing harnessed for user discovery may also reduce the UAV-user interaction, facilitating real-time trajectory design. Furthermore, environment-related sensing results (e.g., locations of scatterers) facilitate seamless FAS adaptation, without requiring CSI estimation. Additionally, the sensing results can be used for constructing a 3D map of the environment, including both dynamic and quasi-static information. This map enables the modeling of wave propagation through ray-tracing techniques, thereby reducing the need for environmental control and ultimately the overhead. Apart from facilitating smart propagation engineering in a single cell, ISAC can also reduce the corresponding time/power consumption of handovers between, for example, UAV-mounted BSs and terrestrial BSs, and thus conserve time and power resources.

\begin{figure}[t]
	\centering
	\includegraphics[width=8.4cm]{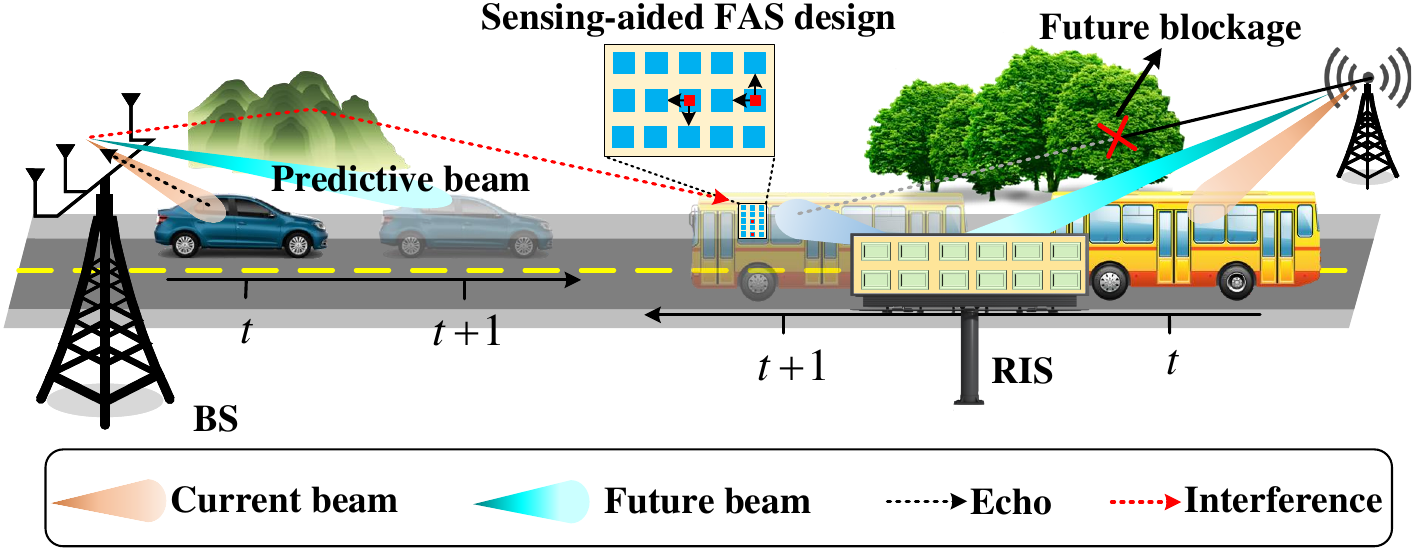}
	\vspace{-3mm}
	\caption{Illustration of ISAC for smart environments.}
	\label{figure3}
\end{figure}

\subsection{Predictive Resource Allocation for Smart Environments}

Based on the estimated user and target dynamics as well as service requirements of S\&C, predictive resource allocation for UAV, RIS, and FAS can reduce service delays and improve reliability. However, it is challenging to jointly optimize the resource allocation of multiple nodes due to the coupled variables. A possible solution is to actively/passively track the state of the served/detected users/targets only individually \cite{Meng2023Sensing}. The system can then combine local sensor data and predict the trajectories of the corresponding users/targets, as shown in Fig.~\ref{figure3}. By jointly optimizing the network's resource allocation and user/target scheduling together with the smart environment, high-quality S\&C services can be achieved. In addition to predictive design based on sensing results, more efficient resource allocation can be supported based on predicting demands for data transmission services. 

\subsection{Blockage Awareness for Channel Control}
While the above potential performance enhancements are promising, they erode due to potential blockages. Hence, blockage awareness through sensing is beneficial for reliable and robust S\&C services in dense obstruction scenarios. According to the estimated movement of users/targets, the UAV location can be optimized based on accurate environmental information to avoid blockages and reduce propagation losses. For near-field S\&C, the antenna or surface array may be partially blocked \cite{singh2023wavefront}, as shown in Fig.~\ref{figure4}(b). In other words, the energy of each user/target is focused on a particular section of the array, as different regions of the array may observe different users. Furthermore, large targets themselves may cause potential blockages, and thus the propagation engineering should be designed by giving cognizance to blockages. In turn, blockages can be exploited for significantly reducing interference between users and cells. As a result, environmental awareness through ISAC offers a breakthrough capability for controlling propagation based on user locations to avoid outages.

\section{Case study: Propagation Engineering for ISAC}
\label{CaseStudy}

\begin{figure}[t]
	\centering
	\includegraphics[width=8cm]{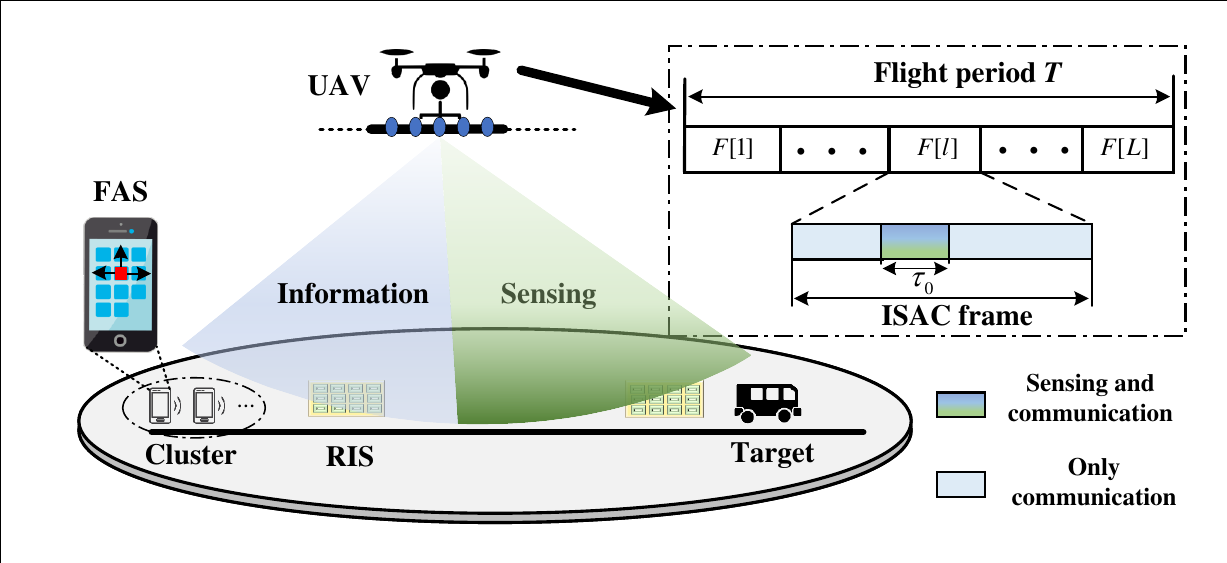}
	\vspace{-3mm}
	\caption{Illustration of simulation scenarios.}
	\label{figure2}
\end{figure}

\begin{figure}[t]
	\centering
	\includegraphics[width=7.5cm]{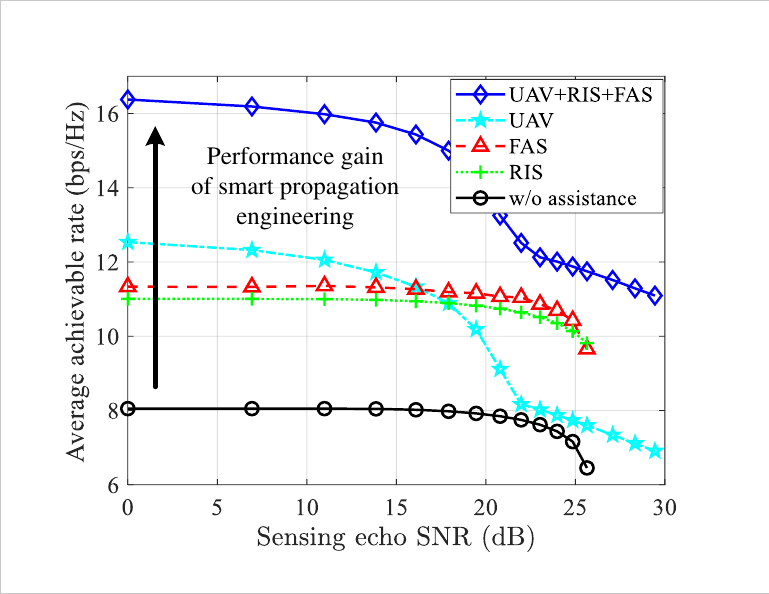}
	\vspace{-3mm}
	\caption{Rate improvement by smart propagation engineering.}
	\label{figure5}
\end{figure}

To demonstrate the efficiency of propagation engineering for ISAC, we consider the optimization of communication performance under sensing performance constraints. As shown in Fig.~\ref{figure2}, we jointly optimize the resources of a UAV, a RIS, and several users supported by a FAS. Fig.~\ref{figure5} characterizes various propagation engineering designs and their achievable rates under different sensing performance constraints. The UAV's maximum horizontal flight speed is set as 30 m/s with a flight altitude of 30 m. Additionally, the noise power at the user and the UAV are set to -70 dBm and -90 dBm, respectively, and the maximum transmit power is 1 W. 

Observe from Fig. \ref{figure5} that collaborative smart propagation engineering using all three techniques significantly enhances the rate attained, compared to the unassisted scenarios. However, as the sensing echo signal-to-noise ratio (SNR) increases, the achievable rate of the communication schemes relying on the assistance of smart propagation engineering decreases rapidly. The primary reason is that smart propagation engineering systems tend to dedicate more resources to the sensing task due to the high sensing SNR required at the cost of communication performance erosion. When the sensing SNR requirement exceeds a certain value, it becomes impractical to achieve satisfactory sensing performance without the assistance of smart propagation engineering. By contrast, the schemes relying on UAV systems can extend the region of adequate sensing performance, while significantly improving communication performance, especially for lower sensing demand. This is because accurate UAV trajectory design significantly reduces the pathloss by performing S\&C tasks closer to targets and users.

\section{Open Problems}
\label{Extensions}
The horizon for research and innovation in both smart propagation engineering and ISAC technologies is wide open. Some of these are discussed as follows.

\subsection{AI assisted Smart Environment for ISAC}  

Artificial Intelligence (AI)-based approaches offer a promising solution for the scenarios of hostile wireless channels and user mobility, eliminating the need for time-consuming iterations routinely harnessed in traditional algorithms \cite{Evmorfos2023Actor}. AI facilitates the real-time adaptation of smart environments by predicting future network states using the recent propagation data gathered. Furthermore, the complexity of the holistic smart environment design across UAV trajectories, RIS phase shifts, and FAS positioning, may be reduced by AI solutions upon evaluating only a fraction of the combined search-space. The sensing functionality of ISAC in turn has a lot to offer for harnessing intelligence in the face of uncertainty. Real-time data gathering by sensing throughout every transmission is the only way to create native network intelligence. Finally, federated learning (FL) provides a promising solution for privacy-preserving distributed training. One could envisage each smart system updating its local AI model based on ISAC data and sending the parameters to a central server for global AI model update in a FL manner. This requires a bespoke training algorithm for the ISAC process, which deserves further investigation.

\subsection{Smart Environments for Secure ISAC}
ISAC comes with its own unique security challenges due to the shared use of spectrum and the broadcast nature of wireless transmission. While smart propagation engineering strikes a flexible balance in simultaneously achieving secure S\&C, the high complexity of obtaining the locations and channels of potential eavesdroppers makes securing ISAC services particularly challenging. Furthermore, sensing imposes vulnerability on the network, since an unauthorized sensing receiver may eavesdrop on the target's location information, or may illegitimately acquire situational awareness. Unlike communication security, which can be designed at the data level to prevent eavesdroppers from obtaining information, sensing security can only be designed at the physical layer. Therefore, how to adapt smart environments to facilitate S\&C security is an issue worth studying. 

\subsection{Smart Environments for Vehicular ISAC Cooperation}
For highly dynamic vehicular networks, developing a general solution for smart propagation engineering has substantial challenges. Considering that vehicles on the same road often have similar mobility characteristics, particularly for platoons travelling in a specific formation, it is advantageous to exploit these similarities for minimizing control overhead by designing vehicular cooperation schemes. Smart environments can help vehicles collaboratively obtain information in a wider field of view, thereby providing more comprehensive situational awareness for transportation. By contrast, the sensing results obtained by vehicular onboard sensors can be used for efficient control of smart environments.

\section{Conclusions}
\label{Conclusions}

Integrating multiple smart propagation engineering techniques with ISAC presents significant design challenges, but also compelling opportunities. We have discussed new design considerations and highlighted essential challenges for the joint design of smart environments and ISAC, revealing their mutual benefits. We further showcased the benefits of these concepts through an ISAC case study. Given the relatively uncharted territory of ISAC empowered by smart propagation engineering, this article endeavored to inspire future research in this field.

\footnotesize  	
\bibliography{mybibfile}
\bibliographystyle{IEEEtran}

	\end{document}